\documentclass[]{article}
\usepackage{stmaryrd,amssymb,amsmath,amsthm,url,supertabular}
\title{Putting Instruction Sequences into Effect}
\author{
	Jan A. Bergstra \thanks{\small Address: Science Park 904, 1098 XH, Amsterdam, 
	The Netherlands. Email: {\tt j.a.bergstra@uva.nl}. This work has been carried out with the support of the NWO Project 
	``Thread Algebra for Strategic Interleaving''. The author thanks Kees Middelburg and Inge Bethke (both UvA) for
	many helpful comments on previous versions of this paper.}\\
	\newline\\
	  Section Theory of Computer Science\\
	  Informatics Institute, Faculty of Science\\
	  University of Amsterdam}
	   
\date{}

\newcommand{\tr}{\ensuremath{T}}
\newcommand{\fa}{\ensuremath{F}}

\theoremstyle{definition}

\begin{document}

\maketitle

\begin{abstract}
An attempt is made to define the concept of execution of an instruction sequence. It is found to be a special case of directly putting
into effect of an instruction sequence. Directly putting into effect of an instruction sequences comprises interpretation as well as execution.
Directly putting into effect is a special case of putting into effect with other special cases classified as indirectly putting into effect. Indirectly putting into effect an instruction sequence
refers to a class of mechanical processes by default, which includes mechanisms making use of preparatory transformations and compilations of
the entire instruction sequence or of significant parts thereof, where the transformations and compilations are themselves produced by means of putting yet other instruction sequences into effect. Returning to instruction sequence execution, its initial informal definition is: directly putting into effect an instruction sequence in such a way that it does not represent a form of interpretation, where a method of directly putting into effect is considered an interpretation if it represents an instance of  the functionality produced by putting some other instruction sequence into effect (either directly or indirectly).
\end{abstract}

\section{Introduction}
\label{sect-introduction}

Recently, issues concerning the following subjects from the theory of
computation have been investigated from the viewpoint that a program is
an instruction sequence:
semantics of programming languages~\cite{BB2007,BM07e,BM07f,BP2009},
expressiveness of programming languages~\cite{BergstraBethke2011,BM07g,BM08h,PZ06a},
computability~\cite{BM09m,BP04a}, and
computational complexity~\cite{BergstraBethke2010,BM08g}.
Performance related matters of instruction sequences have also been
investigated in the spirit of the theory of computation~\cite{BM09i}.
Moreover, in the area of computer architectures, basic techniques aimed
at increasing processor performance have been investigated from that
viewpoint ~\cite{BM06d,BM06b},
and the instruction streams that arise on execution of instruction
sequences~\cite{BM09c} have been studied.
The results of these investigations are evidence of the workability of
the viewpoint that a program is an instruction sequence.

At any rate, instruction sequences constitute a very primitive concept
and execution is what instruction sequences are intended for, at least
originally.
The concept of \emph{instruction sequence execution} (ISE) has been used many times
in the work referred to above, mostly in explanatory remarks, but
without a precise definition that conveys the essence of instruction
sequence execution and is sufficiently computer technology independent.
No such definition is found in other work either, although instruction
sequence execution, however defined, plays a central role in the technology
of most past and current generations of digital computers.

When asked to define instruction sequence execution as a notion in 
a theoretical context this appears to be difficult. Opinions concerning the 
matter diverge and definitional design decisions need to be made.
This paper provides an answer to that question, starting from the description
of a more general notion: {\em putting an instruction sequence into effect} 
(abbreviated as PISiE below).

Having available that notion the task ahead reduces to: obtaining a definition
that explains instruction sequence execution such that (i) it can be
distinguished from methods of putting instruction sequences into effect
that do not qualify as instruction sequence execution and (ii) it also abstracts
from details determined by the ever-changing computer technology.

\subsection{Approach}
Using the general notion of PISiE as an umbrella several more specialized 
versions of it will be determined, notably \emph{directly putting an instruction sequence into effect} (dPISiE), \emph{interpretation},
and \emph{execution}.

The definition of analytic execution architectures given in~\cite{BP04a}
serves as a point of departure.
That definition captures direct putting into effect of instruction
sequences, except for the fact that it must be a mechanical process
by default.%
\footnote{%
PISiE by hand is non-mechanical and the attribute ``by hand'' overrules the
default (of mechanical working) which PISiE carries along.%
}
The working of a so-called analytical execution architecture provides 
an a priori intuition about the
directly putting into effect of instruction sequences.
The intuition concerned suggests that directly putting into effect is either
execution or interpretation.
Directly putting into effect does not encompass methods of putting
instruction sequences into effect that involve compilation. Preprocessing
phases for compilation, verification, or optimization, render a method
of putting into effect indirect.
The central question is what are the conditions under which a method of
putting instruction sequences into effect qualifies as execution,
qualifies as interpretation, et cetera.

Two issues that will complicate the definition of the concept of
instruction sequence execution considerably must be mentioned.
Firstly, informally, execution of an instruction sequence is direct
putting into effect of an instruction sequence in a way that is not made
happen by putting some other instruction sequence into effect.
The negative part of this informal definition causes difficulties,
because it seems to be recursive and somehow not well-founded.
Secondly, instruction sequence execution, as well as several other
concepts that concern methods of putting instruction sequences into
effect, is a concept whose instantiations are expected to be present in
a variety of circumstances, that is technological designs,
and therefore a concept to which a single
definition perhaps does not apply.
This means that, despite its high level of reality, it is possible that
a definitive definition of instruction sequence execution is not
attainable and a definition of a more or less impermanent character must
be opted for.

This open-endedness of the subject
has led to the idea to introduce the concept of the \emph{executionality}
of a method of putting instruction sequences into effect.
Executionality can informally be defined as the degree of being
instruction sequence execution.
Then a method of putting instruction sequences into effect could be
considered to qualify as instruction sequence execution if its
executionality is maximal when compared with other methods of putting
instruction sequences into effect.
This vista will be expanded.

The development of the conceptual framework will include definitions of
the concepts of instruction sequence putting into effect, instruction
sequence execution, and instruction sequence interpretation, definitions
of methods of putting instruction sequences into effect where
instruction sequence execution and/or instruction sequence
interpretation are combined with compilation (such as methods involving
just-in-time compilation, see e.g.~\cite{Ayc03a}), a definition of the
concept of executionality of a method of putting instruction sequence
into effect, and a definition of the concept of executability of an
instruction sequence.

The most closely related work is work in which instruction sequence
execution is considered a relevant concept.
This includes mainly work on
processor design (see e.g.~\cite{HH03a,IYT01a,KHL02a,Sch93a}),
verification of processor design (see e.g.~\cite{CIR00a,Pix96a}),
timing analysis of computer-based systems
(see e.g.~\cite{CLK94a,KP07a,RS09a}),
power analysis of computer-based systems (see e.g.~\cite{FGSS98a}), and
processor performance tuning (see e.g.~\cite{FKLU05a}).
However, in all of it, no serious attempts are made to explain the
concept of instruction sequence execution.

\subsection{Technical aspects of instruction sequences}
\label{sec:1}
This paper is written in terms of instruction sequences, although it might have as well be written in terms of 
computer programs in general, because we have a workable understanding of the former
notion.

Instruction sequences are made up from control instructions, mainly jumps and variations thereof, as well as basic actions which may or may not return a boolean value that decides on how to proceed with a run.   We will assume that  instruction 
sequences are presented
in one of the the notations of \cite{BergstraLoots2002}. Within
an instruction sequence a decision point takes the form of an instruction $+a$ or $-a$ where $a$ is a basic action. The intuition is that in an instruction sequence $X;+a;u;Y$ if $+a$ is executed the action $a$ is performed by both (optionally) changing a state and immediately thereafter
producing a boolean value (the result of $a$); then if $\tr$ is returned execution proceeds with instruction $u$ while if $\fa$ is returned that instruction is
skipped and execution proceeds with the first instruction of $Y$. In $X;-a;u;Y$ execution proceeds with $u$ after a result $\fa$ is produced (as a consequence of processing $a$) and upon the
result $\tr$ execution skips $u$ and proceeds with $Y$. We refer to \cite{Ponse2002} for information on how to extend instruction 
sequence notation with stronger but quite related primitives.
We will call $+a$ and $-a$ steering point 
instructions. In the context of this work we prefer steering point (instruction) to decision point (instruction) in order to provide for a significant distance from 
the terminology of decision making processes and methods. Further we will not use the phrase ``control instructions'' to maintain sufficient distance from the control code terminology of \cite{BergstraMiddelburg2009} which is strongly related to the notion
of dark programming as developed in \cite{Janlert2008}.
The more embracing concept of a program is so hard to define that following \cite{BergstraLoots2002} we prefer to consider 
instruction sequence a more primitive concept.

Suppose an instruction sequence $X$ is given. The basic actions of $X$ are taken from an interface in the sense 
of \cite{BergstraPonse2009}. A simple but yet quite expressive example of a basic action interface is obtained as follows:
assignments to and reading of (effected by steering atoms) boolean variables. Such instructions have been specified in detail in \cite{BergstraBethke2010}. 
We refer to that paper and references cited therein for a specification of syntax and semantics of such instructions. Further examples
of instruction sequences and corresponding interfaces can be found in \cite{BergstraMiddelburg2009b}. \footnote{That paper provides important notations, in particular $X ! H$ for the boolean result of running $X$ over service family $H$, $X \bullet H$ for the effect that running $X$ over service family $H$ has on $H$  (both under the assumption that the basic action interface of $X$ is included in the interface of $H$) 
and $X/H$  for the thread that is produced when $X$ runs over service family $H$ (where it is plausible that the interface of $X$ is only
partially included in the interface of $H$).}

\subsection{Terminology for and around PISiE}
Execution is a prominent member of a family of related notions, each of them applicable with respect to an instruction sequence $X$ on machine $M$. In Table \ref{forms} a listing of terms and phrases is given that can be found in the computing literature all denoting forms of PISiE. The listing is provided without any claim of completeness and without any intention to have orthogonal meanings or to remove different wordings for the same or nearly identical mechanisms. It should convey the relevance of assigning at least informal meanings to a significant number of these terms and phrases so that it can be determined when different terms stand for different content.

\newpage

\bottomcaption{Forms of PISiE\label{forms}}
\tablehead{\hline&\\}
\tabletail{
&\\\hline\multicolumn{2}{|r|}{\small\sl continued on next page}\\
\hline
}
\tablelasttail{&\\\hline}
\begin{supertabular}{|cl|}
01& using $X$\\
02&running $X$\\
03& executing $X$\\
04& execution of $X$\\
05& testing $X$\\
06&interpreting $X$\\
07&interpretation of $X$\\
08&running an executable for $X$\\
09& a (run, terminated run, interrupted run, uninterrupted run) of $X$\\
10&running $X$ until (interrupt, timeout, run-time error, termination)\\
11& $X$ running directly on hardware\\
12& running $X$ on a non-virtual machine\\
13& (automatic, electronic, mechanical) processing of $X$\\
14&(activating, deactivating, activation of, activating a run of, deactivation of, \\
&terminating a run of) $X$\\
15&unintended activation of a (malicious) execution of $X$\\
16& a runtime (for, of) $X$\\
17&executing a runtime for $X$\\
18& running a mobile runtime for $X$\\
19& executing an executable code for $X$\\
20& sequential execution of $X$\\
21&concurrent execution of $X$ (concurrent with an execution of $Y$)\\
22&providing the use of $X$ (as a service, as SaaS, in the cloud)\\
23&(slow, fast, normal, high performance) execution of $X$\\
24&execution of $X$ on (prototype, experimental, dedicated (ASIC)),  hardware\\
25&execution of $X$ in (batch, interactive, debugging) mode\\
26& (correct, valid, invalid, erroneous, faulty) execution of $X$\\
27& unintended execution of (malicious) $X$\\
28& executing $X$ on a virtual machine\\
29&interactive execution of $X$\\
30&speed of execution of $X$ on $M$\\
31& performance of $M$ on a benchmark containing $X$\\
32&a thread taking care of (or: representing, or: constituting) an execution of $X$\\
33& (single-thread, multi-thread) execution of $X$\\
34&operation of $X$\\
35& a (trace, run) of an execution of $X$\\
36& (automated, automatic, mechanical, remote, managed, virtual, non-virtual,\\
& sandboxed, manual) execution of $X$\\
37& monitoring an execution of $X$\\
38& executing a runtime for $X$ in a sandbox\\
39& compilation of $X$ followed by assembly followed by loading 
followed by \\
&execution\\
40&compilation of $X$ followed by execution\\
41& downloading $X$ followed by (execution, interpretation, sandboxed \\
&execution, sandboxed interpretation, managed execution)\\
42& optimizing compilation of $X$ followed by execution\\
43&compilation of $X$ followed by interpretation of the result\\
44&compilation of $X$ followed by an alternating mix of interpretation and \\
&just in time compilation (JIT, that is: further compilation of \\
&fragments which are  executed immediately thereafter)\\
45& as above but with the JIT compiler equipped with a learning subsystem\\
& that takes advantage of information gathered about previous runs\\
&under comparable conditions\\
46&execution of $X$ on a pipelined microprocessor\\
47&pipelined execution of $X$ using micro-threading on a (single core, \\
&multi-core) architecture\\
48&execution  of $X$ by means of a pipelined microprocessor using out of \\
&order instruction issuing (and speculation)\\
49&executing $X$ as a thread on a single core of a multi-core processor \\
&architecture\\
50& executing $X$ as a multi-thread on multiple cores of a multi-core\\
& processor architecture\\
51&simulation (of an execution) of $X$\\
52&a successful test of $X$, (successfully testing $X$)\\
53& an unsuccessful test of $X$ (unsuccessfully testing $X$)\\
54&(successful, unsuccessful, deliberate, intended, unintended, temporary, \\
&experimental)
use of $X$\\
55& formal execution of $X$ on a machine model (execution architecture)\\
56& implementation of a formal execution of $X$ on an execution architecture\\
57& automation of the (formal) execution of $X$ on an execution architecture\\
\end{supertabular}

\subsection{Looking for definitions}
It is a reasonable expectation that instruction sequence theory provides definitions 
of all notions mentioned above. That may  include
 rejecting some of these terms or phrases as redundant, or ill-formed, or confusing. 
 Finding definitions for all items in the list is too much of an effort but the following 
 items must be included: execution, run, interpretation, JIT based interpretation, 
 pipelined execution (including advanced features: micro-threading, out-of-order 
 issuing, speculation).
 
 Although it seems plausible to begin with a definition of ISE
 and to build all further definitions on top of that concept, we found reasonable evidence 
 that this strategy will not work. Taking interpretation as the point of departure is not satisfactory
 either. Instead we take PISiE and dPISiE  as points of departure. 
 Once these concepts are in place progress can be made towards defining most other 
 mentioned terms and phrases.
 
 Besides choosing a strategy for the order in which definitions are to be developed
 there is the rather more complex and even philosophical issue concerning 
 what kind of definitions  are to be given and what has to be expected from a definition
 once it has been agreed upon. These matters will be briefly dealt with in the 
 concluding remarks only.

\subsection{Some requirements on usage mechanism meanings}
In Table~\ref{prop} some properties are listed that we consider relevant for notions of instruction 
sequence usage including PISiE and dPISiE. We even hold that these properties may be 
considered to constitute requirements on the definitions of dPISiE end PISiE.

As far as we can judge, the properties  in Table~\ref{prop} will indeed result once the definitions 
of PISiE and dPISiE have been completed. 
Thus, assuming that dPISiE is a subclass of PISiE and that 
compilation stands in the way of directness, these items may be understood
as constituting a specification of what we want to achieve in terms of 
organization of concepts.

\mbox{}\\
\bottomcaption{Requirements on PISiE and dPISiE of $X$.\label{prop}}
\tablehead{\hline &\\}
\tabletail{
&\\\hline\multicolumn{2}{|r|}{\small\sl continued on next page}\\
\hline
}
\tabletail{&\\\hline}
\begin{supertabular}{|ll|}
01&ISE of $X$ is an instance of dPISiE of $X$.\\[0.1 cm]
02&dPISiE of $X$ is always an instance of PISiE of $X$.\\[0.1 cm]
03 &Interpretation of $X$ is an instance of dPISiE of $X$.\\[0.1 cm]
04& If compilation is part of the method a mechanism is not classified as\\
& dPISiE (thus compilation renders a method of putting into effect indirect).\\[0.1 cm]
05&Compilation of $X$ followed by execution of the result is an instance of \\
&PISiE of $X$ but not of dPISiE of $X$.\\[0.1 cm]
06&An interpretation of $X$ is not an execution of $X$.\\[0.1 cm]
07&An interpretation of $X$ is an execution of $X$ on a virtual machine but a\\
& virtual machine is not a machine.\\[0.1 cm]
08&A run of an execution of $X$ is a run of $X$ (but not conversely).\\[0.1 cm]
09&A manual execution of $X$ is not an execution of $X$ \\
&(just as a crashed car is not a car).\\[0.1 cm]
10&A simulation of an execution of $X$ is not an execution of $X$.\\[0.1 cm]
11&Execution of $X$ is a concept that evolves together with the evolution of \\
&processor technology.\\[0.1 cm]
12&Qualifying PISiE as an execution depends on the working of a particular \\
&processor architecture. Execution cannot be defined uniformly in\\
& mechanical terms.\\[0.1 cm]
13&Execution is about how  a run works  in mechanical terms, not about what \\
&is achieved in terms of performance. So in some architectures an inter-\\
&pretation of $X$ may be faster than any execution of $X$ (at least in\\ 
&principle).\\[0.1 cm]
14&Testing $X$ need not involve execution of $X$.\\[0.1 cm]
15&Testing $X$ involves PISiE of $X$.\\[0.1 cm]
16&Using $X$ involves PISiE of $X$.\\[0.1 cm]
17&Running $X$ in debugging mode is not an instance of ISE.\\[0.1 cm]
18& Providing the use of $X$ as SaaS may or may not involve ISE of $X$.\\[0.1 cm]
19&Executing $X$ on dedicated hardware (ASIC) $H$ is a misnomer if $H$ \\
&is dedicated for producing the functionality produced by executing $X$ \\
&(on another platform). It is a valid notion if $H$ works in a generic/uniform \\
&way for many instruction sequences with a comparable instruction interface \\
&and a comparable (or lower) length.\\
\end{supertabular}

\subsection{The mechanical default}\label{Prequirements}
It is necessary to resolve the potential ambiguity between formal execution (execution as a concept in theory) and real execution (execution as an instance of  real machine behavior). 
 The following assumptions can be understood as restrictions on the concept of an execution which are imposed before a satisfactory definition is 
provided.\footnote{This way of doing is well-known from axiomatic methods in mathematics and logic, where readers may be asked to 
understand a sequence of axioms about a concept which becomes gradually known by capturing the combined meaning of 
the full list of axioms.}
\begin{enumerate}
\item ``Real execution'' refers to the physical process of running an instruction sequence on a 
machine rather than to its formal 
counterpart which is referred to as ``formal execution''.
\item ``Execution'' for short is used as a short-hand for ``real execution'', while the adverb 
``formal" is needed when referring to its 
theoretical counterpart.\footnote{Hence a formal execution is not an execution.}
\item Execution of an instruction sequence takes place on a particular machine at a particular place, during a particular interval of 
time and triggered by a particular sequence of preparatory events. It can take place at different machines, geographical positions, time intervals. ``Execution'' refers to a particular execution in principle.
\item It is possible to speak about a past, a present or a future execution of $X$ on machine $M$ of machine type $T_M$.
\item For a given machine type $T_M$ different executions of a given instruction sequence $X$ on machines of that type have many 
aspects in common. Thus $T_M$ executions of $X$ constitute a subtype of executions of $X$. This subtype includes future, or
rather conceivable, executions as well.
\end{enumerate}

More generally all forms of PISiE will assumed to be mechanical processes unless 
stated otherwise.

\subsection{Retrospective definition of instruction sequence execution is inadequate}
A \emph{retrospective} definition\footnote{A retrospective  definition may alternatively be called a {\em reverse engineered definition}.}  of ISE  is an explanatory text which can be used to make sense of previous usage of the concept in work on the theory of instruction sequences in the style of program algebra. By nature a retrospective definition is ad hoc, because its deficiencies 
cannot be repaired without compromising the major objective that such definitions provide additional explanation for significant
previous usage.

Retrospective definitions can occur in all circumstances where one is dissatisfied with the quality of definitions used for a concept in some body of literature, while intending not to rewrite that work. There is no intrinsic property of the subject which makes it likely to be provided with a retrospective definition. Different bodies of work making use of the same concept may lead to different retrospective definitions. In some cases the use of a term shows irreconcilable discrepancies between some texts making use of it. That indicates an inconsistency for which a resolution may be needed when further work is to be carried out. Resolving an inconsistency can make some work obsolete in favor of other work that remains up to date.

We will now look in more detail into retrospective definition of ISE. We have already noticed  that the phrase instruction sequence execution has been made use of many times  in previous work of our colleagues and ourself without creating a pressing need for a more precise definition. A definition that covers most usage is as follows:

\begin{quote}
For an instruction sequence $X$ written in notation $L$ and a service family $H$ a {\em manual execution} of $X$ over $H$ 
amounts to a manual production of a progression as specified by the operational semantics of $X$ in the context of $H$.

An instruction sequence $X$ is {\em automatically executed} if some machine $M$ automatically performs the 
basic actions, steering atoms and jumps of
$X$ until termination. It is assumed that the machine has been designed specifically for (or rather with among its most prominent tasks) automating the manual
execution of instruction sequences of notation $L$ over a basic action interface that includes the basic action
interface of $X$ in execution architectures
with a service family that has the same interface as $H$.

Execution of $X$ either refers to a theoretical (formal) description of an automatic execution of $X$ on some appropriate 
machine $M$, often with time and space coordinates left unspecified,
or to an automatic execution on a physical machine which actually  takes place or has taken place in space and time. The choice
between these meanings is to be found from the context of usage of the term ``execution".
\end{quote}

We consider this particular retrospective definition of ISE  to be unsatisfactory for the following reasons:
\begin{enumerate}
\item The definition depends on three further notions which are left entirely without an explanation: 
\begin{itemize}
\item manual execution of an instruction sequence, 
\item automation of a manual execution by means of a machine, and
\item a machine's design being dedicated to precisely this form of automation, or more precisely to have this form of automation 
among its primary tasks.
\end{itemize}
In this way execution is not convincingly reduced to simpler notions.
\item There is no indication that execution of an instruction sequence might be qualitatively 
different from its interpretation (that is execution of an interpreter, itself represented as an instruction sequence, 
with $X$ as an input), or from a simulation of $X$ (or rather from a simulation of an execution of $X$). So the definition
fails to isolate execution properly  from neighboring notions within a range of relevant comparable notions.\footnote{The required (but missing) intuition may be that execution takes place directly on the hardware without any mediation by other processes involving
instruction sequence execution. But these additional comments do not qualify as a definition in the absence of a clear story about the distinction between software and hardware which is an important but different issue.}
\end{enumerate}

\section{Putting an instruction sequence into effect}
An instruction sequence usually contains precise information on how to perform a specific task.  However, an abstract description of that task
is not easily derived from an instruction sequence. Given an instruction sequence the corresponding task is an implicitly assumed 
mental construction
introduced by human observers that helps to make sense of the instruction sequence: why does it exist, why has it been designed 
the way it is, who cares about its existence. When providing definitions there is no access to this implicit abstract task even if it exists in the minds of one or more persons involved.

Let $X$ be an instruction sequence. With {\em putting $X$ into effect}  we mean any
physical process that follows the prescriptions of $X$ in such a way that all or part of its underlying ``task'' are achieved. 
 The product of putting $X$ into effect is a \emph{progression}, that
is a sequence of events. Another word for a progression is a run. A run can be hand made or computer generated.

The mechanical defaults imposed for ISE in Section \ref{Prequirements} 
above are all to be required from 
putting into effect as well. Then nothing more is said than that PISiE is a process that takes place in space and time by means of specific technology. 
\subsection{Directly and indirectly putting into effect}
The terminology of putting into effect suggests two closely  related notions:
\begin{description}
\item{{\em Directly putting into effect.}} A mechanism for putting instruction sequences into effect is said to work in a direct fashion if it operates in the order of the operational semantics without any significant amounts of (pre)processing taking place in advance or, at irregular times during the production of a progression. Interpretation of an instruction sequence by a machine by way of putting another instruction sequence (the interpreter) into effect is considered direct. Execution, however defined, is also assumed to be a special case of directly putting into effect. An appropriate imaginative definition of dPISiE is found in previous work on execution architectures 
(see \cite{BP04a}).\footnote{Running $X$ means exactly the same as directly putting $X$ into effect.}
\item{{\em Indirectly putting into effect.}} If a mechanism for putting instruction sequences into effect makes use of preparatory stages such as compilation, optimization, dead instruction removal, it is said to be indirect. It may be that the second phase of a method of indirectly putting $X$ into effect involves the direct putting into effect of $Y$, which has been obtained by transforming $X$. Then still, as a method of putting $X$ into effect the whole process is considered indirect.
\end{description}

\subsection{Directly putting into effect defined in more detail}
A positive formal definition is obtained by stating that a progression produced by an instruction sequence in an execution architecture as specified in \cite{BP04a} constitutes an instance of dPISiE. 

We recall that an execution architecture contains a number of local services, whose state is initialized to a fixed initial state in advance of each new run. Instruction sequences contain basic actions taken from an instruction interface. Target actions are meant to be applied to external services. The performance of target actions in the right order is the very purpose of PISiE. Co-target actions are processed by local services. The only reason for the use of co-target actions of $X$  is to enable the machine putting $X$ into effect to perform its target actions in the right order.

dPISiE allows changing the order of use of co-target actions as long as this has no semantic effect visible at the level of co-target actions. A basic action interface may provide information about the boolean result that basic actions will return and this information may be made use of when reordering the natural order of co-target actions as given by an instruction sequence. In the case 
of directly putting into effect it is required that the ordering (including simultaneous invocation for reasons of processing speed) of co-target actions is computed between target actions rather than in an optimization phase in advance and that the computation which determines the order is done without putting another instruction sequence 
into effect.\footnote{Micro-programming may be used, but because that makes use of firmware it lacks the flexibility required from a method of putting instruction sequences into effect.}

\subsection{Related notions}
Once PISiE has been established as a concept, several related questions can be posed and answered.
These notions are not strictly related to PISiE, but need to be contemplated in order to have firm ground 
needed to experiment with the concept of execution. 
\begin{description}
\item{{\em Who puts into effect}?} When putting $X$ into effect the question arises who is responsible of 
initiating that activity. Implicitly there is an operator around who plays that role. 
Indeed ``putting into effect" has two arguments (who puts and what is put) but
running can be used with a single argument (what is running) or with two arguments (who is running what). If one says that $X$ is running on a machine then the corresponding question who is running $X$ is less pressing because one may choose the 
single argument use of the verb running when reading this sentence.

\item{{\em Are the effects real and intended}?} If $X$ is put into effect then the effects may deviate from what was intended when 
$X$ was created. This typically happens if the resulting progression is considered a test, or if it serves an educational purpose
for prospective users of the same instruction sequence in so-called {\em realistic} conditions. It can also happen if a construction error has
occurred when producing $X$ or if an error of analysis generated unjustified expectations.

\item{{\em On the fly existence of instruction sequences.}} A program can be put into effect by compiling it into an instruction sequence and thereafter putting that instruction sequence into effect. For instance if the instruction sequence is executed 
(rather than merely 
put into effect) one also speaks of an execution of the program in its original and uncompiled form. In the case of 
JIT compilation only fragments of an instruction sequence are determined and put into effect. Running a program by 
means of JIT compilation implies running an instruction sequence that exists on the fly (and for that reason at any moment partially) only.
\end{description}

\section{Interpretation, execution, and compilation}
We are  now able to provide further definitions:
\begin{description}
\item {\em Interpretation.} The putting into effect of instruction sequence $X$ qualifies as an interpretation if it is producing a progression $\pi$ such that:
\begin{itemize}
\item $\pi$ qualifies as directly putting into effect of $X$, and,
\item $\pi$ is itself the effect (result%
\footnote{%
Remarkably it is impossible to find any further specification or definition, at this
level of abstraction, of what it means for $\pi$ to be the result of another putting into effect.%
}%
) of the putting into effect of another instruction sequence $Y$%
\footnote{This definition allows that $Y$ itself is interpreted, or compiled and subsequently interpreted, which calls for some instruction sequence $Y^{\prime}$ acting as an interpreter and so on. But it must be well-founded, at some an instance of stage directly putting into effect must be found.}
 on the same piece of equipment, and,
\item the latter is uniformly the case in the sense that for a large family of alternative instruction sequences $Z$ putting $Y$ into effect to data which contain a representation of $Z$ 
results in a progression $\pi\prime$ constituting  the directly putting into effect of $Z$.
\end{itemize}

\item{\em Execution in the case of single thread systems.}  On machines disallowing any form of multi-threading and concurrency the following definition of instruction sequence execution is adequate. A progression $\pi$ constitutes an execution of $X$ if :
\begin{itemize}
\item $\pi$ is an instance of putting $X$ into effect, and, 
\item  for no instruction sequence $Y$, $\pi$ is the result%
\footnote{This result is the primary result of putting $Y$ into effect, and obtaining that result constitutes the reason for putting $Y$ into effect. All other results or computational outcomes of PISiE for $Y$ are secondary, and may be considered side effects.}%
 of  putting $Y$ into effect to data that represent $X$, and,
\item $\pi$ is an instance of directly putting $X$ into effect.
\end{itemize}
\item{\em Executionality based definition of execution.} Often in a machine different instruction sequences are simultaneously being put into effect in various different ways. In these more general circumstances 
execution requires a more liberal definition.

Then a {\em ``degree of executionality''} is assigned to each of these puttings into effect and those instruction sequences are said to be executed for which the degree of executionality is considered positive. The degree of executionality ranges between $-1$ and $+1$. Using this terminology, a progression constitutes an execution of $X$ if:
\begin{itemize}
\item it is an instance of putting $X$ into effect, and, more specifically
\item it is an instance of directly putting $X$ into effect, and,
\item in the context of the machine $M$ realizing the putting into effect of $X$ the latter process has positive executionality.
\end{itemize}

\item{\em Executability relative to architecture $A_m$.}
An instruction sequence $X$ is executable relative to machine architecture  $A_m$ if  any 
machine of that architecture can be prepared into a state from which it executes $X$. 

\item {\em Compilation into object code.} A putting into effect of $X$ is based on compilation into object code if it comprises the following phases:
\begin{itemize}
\item transforming data representing $X$ into a representation of an instruction sequence $X^{\prime}$ by means of putting into effect an instruction sequence $C$, and subsequently,
\item directly putting $X^{\prime}$ into effect, by way of execution.
\end{itemize}
\item {\em Compilation into intermediate code.} A putting into effect of $X$ is based on compilation into intermediate code if it comprises the following phases:
\begin{itemize}
\item transforming data representing $X$ into a representation of an instruction sequence $X^{\prime}$ by means of putting into effect an instruction sequence $C$, and subsequently,
\item directly putting $X^{\prime}$ into effect by way of interpretation, or,
\item putting $X^{\prime}$ into effect by means of compilation into object code.
\end{itemize}
 
\end{description}

Most of the notions listed in Table~\ref{prop} as forms of usage mechanisms can now be provided with definitions on the basis of the ones just developed. 

\subsection{Result sandwich form definitions}
Instruction sequence execution is defined as an instance process of type $F$ where the process $F$ itself is not at the same time the result of a another process of type $F$, with $F$ standing for directly putting an instruction sequence into effect. This definition has a particular form which arises in other contexts as well but usually without the resulting concept, or defined category, being given a proper name. 

In somewhat formal terms we look at a description of definition of the form:
$G = F \wedge \neg R_p(F)$
where  $R_p(F)$ stands for ``the primary result of $F$''. We will call this the result sandwich form, or in full detail, the primary result sandwich form.

A related form of definition is $G = F \wedge \neg C(F)$ where
$C(f)$ stands for ``caused by $F$''. This may be called the casual sandwich form of definition. Here are some examples:

\begin{itemize}
\item (A primary tumor is) an instance of a tumor growth process which is not caused by another instance of a tumor growth process.
\item A forest fire which is not cased by another forest fire.
\item A traffic accident which is not cased by another traffic accident.
\item An instance of decision taking which is not the result of another instance of decision taking.
\end{itemize}

We conclude that instruction sequence execution can be defined on the basis of PISiE by way of a result sandwich form definition, in conjunction with the additional requirement  that it constitutes and instance of dPISiE.
\subsection{Well-foundedness of PISiE}
The following form of well-foundedness may be claimed: 
\begin{itemize}
\item if machine $M$ is putting instruction sequence $X$ into effect then some (perhaps different) instruction sequence $Y$  is being put directly into effect in $M$, which moreover causes (or equals) the putting into effect of $X$, and, moreover,
\item some (perhaps different) instruction sequence $Y^{\prime}$  is being executed, which directly or indirectly causes the putting into effect of $X$.
\item if a machine is concurrently putting a number  of instruction sequences into effect then at least one of these is assessed with positive executionality.
\end{itemize}

We consider the executionality based type of definition to be most promising due to its potential flexibility. For that reason, executionality, or rather the degree of executionality is in need of further explanation.

Before a metric can be contemplated which measures the executionality of a method for running an instruction sequence a survey is
needed of aspects which must be taken into account when making an assessment of the executionality  of a progression produced by an automated instruction sequence running method. This, however is a major difficulty because there seems
not to be any obvious limitation to these aspects. When technology evolves aspects relevant for executionality may be added and removed.\footnote{For instance we do not provide any connection between executionality and energy consumption in this paper but that does not imply the definitive absence of such a connection. 

Another aspect left aside in this paper concerns the fact that a real machine can (potentially) run only finitely many different instruction sequences, whatever method of running it may be using.
For that reason each machine architecture contributes to the positive executionality of runs of only finitely many instruction sequences. This suggests a fundamental difficulty for defining positive executionality in a machine (architecture)  independent way.}

\subsubsection{Indications for positive executionality}
There are indications for positive executionality, and such indications are conclusive in the absence of overruling negative indications. We will first list positive indications by means of clauses that lead to the expectation of positive executionality. In the absence of indications against executionality (which are listed below in \ref{Against}) the expectation of positive executionality promotes into the judgement of positive executionality.\footnote{Both listings of indications in favor of and against positive executionality may be incomplete in the sense that further investigation suggests additional items that should be added. There is no method available to remedy this problem conclusively.}

Manual execution has lower executionality than all forms of automated execution; simulation is less executional than interpretation. 
Different forms of manual execution may differ in executionality because of differences in the way basic actions are performed.

\begin{enumerate}
\item We  will speak of \emph{positive executionality} of a run of $X$  if it reasonable to consider the run an execution. Positive executionality of $X$ does not exclude that other concurrent runs of $X$ are considered to feature a higher 
degree of executionality.

\item Executionality comes in as the dominant notion which needs to be analyzed in order to obtain a flexible definition  of instruction sequence execution.
\item Positive executionality is expected if a machine executes the normal execution cycle, fetch, decode, execute 
(of an instruction), store. Pipelined versions which produce that cycle more efficiently are also considered to deliver positive executionality.
(A detailed specification of in-order pipelined execution architecture, including a perspective on pipelining providing a 
concurrent version of the normal execution cycle, for instruction sequences can be found 
in \cite{BergstraMiddelburg2008}.)\footnote{Following \cite{BergstraMiddelburg2007,BergstraMiddelburg2009} 
basic actions of an instruction sequence $X$ are to be processed by
services. These services can be distinguished in target services and para-target (or auxiliary) services. Basic actions 
directed to para-target services (so-called para-target actions) are performed in order to ensure that a run performs the 
basic actions meant for target services (so-called target action)  in
the right order (that is the order in which they occur in the thread extracted from $X$). Out of order execution allows
run-time determination of permutations between
different para-target actions and between para-target actions and target actions as long as the ordering of target actions is kept correct.}

\item Beyond pipelining there is multiple pipelining which supports multi-(micro) thread execution. The details of that way of running instruction sequences have been worked out in \cite{BM06d}, where it was found that this is far from obvious. Such 
mechanisms may increase execution speed and decrease the degree of executionality because of an inherent lack of flexibility which necessitates a preprocessing stage which may fail to deliver useful results in a majority of cases.

When considering increasingly complex microprocessor architectures, execution quality increases as long as universally applicable features are added which have acquired a firm footing in practice. Such features may improve speed, reduce clock speed and circuit depreciation, reduce power consumption 
or simplify memory usage to mention some prominent objectives of progress in microprocessor technology. An example of additional complexity is speculation which allows statistically validated guesswork on the outcome of conditional evaluation during a run. This in turn allows to fetch instructions earlier on, though with the risk that the fetch needs to be undone when the guess turned out to be wrong. 

\item Run-time preprocessing of parts of an instruction sequence (or rather of a cached subset of a running instruction sequence)  should not
be considered to decrease executional quality. It need not increase it either. Speed differences may exist between runs of equal (or at least comparable) executional quality.

\item Positive executionality indicates a subset of runs which is likely not to contain the very slow ones but which
 may also miss some or even all very fast ones. Nevertheless, if the class of machines considered consists of actually existing pieces of hardware those with positive executionality are relatively fast as automatons running instruction sequences (taken from the domain of instruction sequences which they can run at all).
 Execution of $X$ need not be the fastest way of putting $X$ directly into effect.\footnote{Indeed it cannot
be excluded that in the future all executions will be slow in comparison to optimized interpretations. Stated differently, machine evolution may render instruction sequence execution obsolete for the majority of instruction sequences.}
\end{enumerate}

\subsubsection{Indications for negative executionality}\label{Against}
Indications against executionality (or for \emph{negative executionality}) are counted more strongly than indications in favor.
\begin{enumerate}
\item An indication of negative executional quality is found if an instruction sequence is not in control of the machine on which it is running.
This happens if another instruction sequence for instance the encoding of an interpreter has a higher claim to positive executional status. This negative criterion need not be invoked in case of  multi-threading because the instruction 
sequences of different threads can be running with equal degrees of executionality.

However, if $X$ is running concurrently with another thread
the  executionality of that run is lower than it would have been in the absence of the second thread. Thus multi-threading 
reduces the executionality of each of the threads, though the subsystem consisting of all threads which are concurrently active may still qualify for positive executionality.

\item A negative degree of executionality covers improved forms of interpretation, for instance interpretation with just in time compilation, and so-called managed execution. Both forms of running an instruction sequence have lower executionality 
than the classical
four stage instruction processing cycle and its pipelined versions. The argument is that the execution manager itself is executed with higher executionality.\footnote{At first sight one may wish to have managed execution subsumed under execution. That need not be the
case, however. Suppose agent $A$  is listing $B$'s properties and finds out that item $i$ has been stolen (from $C$). Then $i$ is a stolen property but not a property of $B$ but rather from $C$.}

\item Positive executionality of a run requires parametrization of the underlying technology: hardware specifically designed to run a particular instruction sequence is not executing that execution sequence and it may have very low executionality. This implies that speed is not the decisive characteristic
which determines the executional quality of a method for running an execution sequence. Dedicated equipment that has an instruction sequence $X$ encoded in its hardware may be running $X$ while not executing it due to lack of flexibility.

\item A machine $M$ that can execute $X$ (that is produce a run with positive executionality) can also execute $Y$ with $Y$ an instruction sequence written in some notation used for designing $X$ and under the constraint that its size is bounded by engineering parameters intrinsic to $M$. If this fails to apply there is a decisive indication against executionality.

\item Preprocessing of an instruction sequence to allow out of order execution may be considered a part of running $X$ but is not included as a part of its execution. That is, it markedly decreases executional quality by including unrelated tasks.

\item If machine $M$ (which is supposed to execute $X$) is included in a large piece of equipment which severely constrains 
the number and size of instruction sequences that $M$ may be made to run, it can only be said to execute $X$ if a modification of
the surrounding equipment can be easily imagined such that the flexibility required in the previous points is regained.
\end{enumerate}

\section{Further matters}
The definitions of dPISiE, PISiE, interpretation, and execution need to be checked in many ways. That cannot be conclusively  performed in this paper. One way to investigate the plausibility of the definitions is to consider related questions and notions and to see to what extent a plausible story results. In this Section we consider the concept of simulation, we consider the 
question that executions may not exist, we look into the notion of an excutable, and we consider instruction sequence testing.
\begin{description}
\item{{\em Can a run be simulated}?} Simulation is generating a run in more abstract 
conditions.\footnote{This usage of simulated execution can be found for instance in \cite{Beizer1970}. According 
to \cite{ForceAustin1998} testing can also be simulated.}
Abstract interpretation often referred to as partial evaluation
is a quite similar notion. Simulation of $X$ is putting $X$ partially into effect (only a part of the effects is taken care of). Thus one 
may hold that every run is a simulation but that not every simulation is a run. Unfortunately in the world of software engineering simulation also refers to a stepwise form of running, a process which is sometimes referred to as debugging (even in the absence
of bugs that might be found or repaired).\footnote{Debugging (see \cite{ArakiFurukawaCheng1991}) is a generalization of testing. While testing is about the determination of errors 
(discrepancies between specified and observed behavior), debugging proceeds to the location of causes (which are assumed to
reside inside an instruction sequence and is identified as a set of subsequences of the instruction sequence) their labeling as bugs and repair of the instruction sequence by means of local modification which replaces the bug by an improved set of instruction sequences.}

\item{{\em Executions of $X$ may not exist.}} As ``execution of $X$'' refers to a subclass of ``putting $X$ into effect'', 
that subclass may
be empty. This happens for instance 
if no known or envisaged automated technology bears a prospect of running $X$ in such a way that
labeling the run as an execution is justified. If $X$ is known as a mathematical object, far to long to ever be stored on a machine
made from particles available in the entire universe, one may assume that executions of $X$ cannot exist.

\item{{\em Executables and executable instruction sequences.}} An instruction sequence is executable if it can be executed on some actual machine or else (if non-existing) on a technically conceivable machine, which is the same as it being an executable control code for some
machine. These observations don't provide a definition but merely a tautology. It is unproblematic to understand an instruction sequence 
as a control code. If executability of a control code can be defined in a productive way that provides a significant candidate for a definition
of instruction sequence executability (i.e. being an executable control code). It is at first sight unclear where this line of thought can bring us.
\item{{\em Executable control codes.}} A meticulous attempt to be conceptually clear about control codes and control code executability has been made in \cite{BergstraMiddelburg2009}.
Following \cite{BergstraMiddelburg2009}
code controlled machines come with a notion of an executable (specific for the machine). However, as is concluded in \cite{BergstraMiddelburg2009}, executability (of a control code) cannot be derived
from the underlying concepts and the extension of the set of executables for a machine is a matter of rather arbitrary definition
even when the technical details of the machine are known. In \cite{BergstraMiddelburg2009} a detailed explanation is given of 
when and why a control code will count as an instruction sequence. In summary, an executable instruction sequence is an
instruction sequence which at the same time is an executable control code for a realistic machine (that is either existing or practically
conceivable). That notion of an executable must be based on a model of an architecture.
It is plausible to assume that the extension of executables grows in time, that is with the improvement of technology
instruction sequences may become executable, while the demise of machinery turns them into museum pieces which keep the
status of being technically conceivable.
\item{{\em Instruction sequence testing.}} A seemingly major use of the concept of execution is in the definition of instruction 
sequence testing: testing is obtaining information about an instruction sequence by executing it. However, given the conventions just 
mentioned it is be acceptable to state that testing is obtaining 
information about an instruction sequence by putting it into effect.\footnote{In \cite{Middelburg2010b} it has been established that 
the software testing literature fails to be explicit about the physical experimentation
behind software testing. Viewing the meaning of an instruction sequence as a function between appropriate mathematical domains, the intermediate result of instruction sequence testing is a finite subset of the graph of that function and ``testing'' as a methodology of 
instruction sequence engineering refers to the process of obtaining information about an instruction
sequence from such intermediate results. ``Testing'' also involves heuristics about  which part of the graph should be created for 
subsequent inspection, given the type of information the tester intends to acquire.}As a consequence the concept of execution can be 
eliminated from a definition of (instruction sequence) testing. One finds that putting into effect is a far more prominent concept than comes about from its introduction as a more easily defined container concept for execution.

The question remains to what extent execution is used for the definition of other concepts concerning instruction sequence processing. The centrality of ``execution'' is an open matter, and we believe that it is a rather isolated notion in the sense that few other concepts are
dependent on its proper definition.\footnote{This observation may explain the lack of definitions of ``execution'' in the 
literature on computer programming.}

\end{description}

\subsection{Technology independence and technology scope}
The definition of directly putting into effect can be blamed for being uninformative from a mechanical point of view. Refinements can be designed by incorporating the conventional five stage fetch, decode, execute, store, and retract 
execution cycle into the definition.
One may proceed by incorporating a pipeline and this pipeline may be made increasingly more complex (up to 30 stages are used in commercial processors). Then micro-threaded pipelining may be taken into consideration. Working this way, however, a definition may rapidly degrade into a kind of photographic picture of a very specific stage  of technological development  thus defeating the intended generality.

The virtue of the auxiliary notion of executionality is that it allows one to accommodate these technicalities in the framework of the existing definition of directly putting into effect as long as positive executionality can be maintained.

Besides future technological development also current techniques can be taken into account in a more explicit way.

\subsubsection{Paging: changing into a broader perspective}
Implicit in the definition of instruction sequence execution is that an instruction sequence has been 
loaded before execution. In fact it can only be the loaded form of the instruction sequence $X$  that is executed, 
and certainly not its mathematical abstraction residing in some 
algebra. If $X$ is too large to fit in fast memory, paging will be used to move parts of $X$ from slower memory to faster memory
by need. The actual process of memory swapping depends on the actual execution. If paging is effected by the hardware 
its presence does not stand in the way of high executionality. If, however, paging is done by means of executing another instruction sequence (that is, paging is code controlled) executionality decreases. If paging takes place every 2 or 3 steps on average executionality has become negative, if on average a page swap takes place after hundreds of steps executionality may be considered positive even if the paging mechanism results from execution of an instruction sequence.

Summing up, paging introduces a broader perspective on instruction sequence execution, which by default is a different perspective as well.

\subsubsection{Instruction sequence expressions, a further change of perspective}
In \cite{BergstraMiddelburg2009b} an {\em instruction sequence calculus} is discussed with generalized text sequential composition.
This leads to instruction sequence expressions which can be rewritten into instruction sequences by expanding 
subexpressions making use of bound instruction sequence variables to instruction sequence expressions containing no such variables. Logically an instruction sequence expression (hereafter {\em inseqex}) represents an instruction sequence and for that reason an inseqex may be considered an instruction sequence itself. Contemplating the execution
of an inseqex $X$ it can be concluded that this requires the evaluation (a better term might be expansion) of $X$ into an 
instruction sequence $Y$ not involving bound instruction sequence 
variables which subsequently must be loaded so that it can be executed, with or without paging mechanism. Expansion requires a preprocessing stage most likely controlled by a putting another instruction sequence into effect for that purpose. This leads to indirectly putting the inseqex into effect.

It is easy to find examples where the evaluation of $X$ requires a combinatorial explosion in time and space which makes it unfeasible. In such cases indirectly putting $X$ into effect cannot be done by means compilation (that is: expansion) immediately followed by  execution even if 
code controlled paging is allowed. It follows that interpretation is the only option. This raises an interesting question: is the expressive power of inseqex's perhaps 
strictly larger than the expressive power of instruction sequences for the same machine? Such questions need to take
bounds of the size of loaded instruction sequences into account. If that question has a positive answer
it may provide an explanation of why modern program notations
tend towards being interpreted. It also leads to the subsequent question as to whether or not an interpreter for instruction sequences
can be designed itself as an instruction sequence (i.e. disallowing generalized text sequential composition). 
If not, a second level of interpretation is needed.

It is plausible to assume that an inseqex can be transformed in a piecewise fashion into ``localized'' instruction sequences (instruction sequence fragments) which can be executed until a jump outside a fragment is made and a new instruction sequence fragment surrounding the goal of the jump needs to be generated, which is the executed. In this way a run oscillates between execution (of a localized instruction sequence as long as control stays within), interpretation (first and last steps of the run of an instruction sequence fragment)  and compilation for creating instruction sequence fragments.

\subsubsection{Conjectural technology}
One needs to contemplate the role of technology in providing definitions of directly putting an instruction sequence into effect for
increasingly long instruction sequences and for increasingly complex basic action interfaces. 
Definitions can be made more demanding in mechanical terms thus ruling out successive generations of computer technology. Can a highly demanding 
definition be given (perhaps of a special case of directly putting into effect) even if no technology is available that complies with the definition? We suggest that hypothetical technology may be used to
define future executions, even in cases where the class of past and current execution is still empty, 
and no future execution is envisaged on a 
machine that is currently in existence. In such a case the definition can be given, while no examples of it can yet be 
established.\footnote{This might imply the rather implausible situation that the notion of instruction sequence execution
specialized to a specific instruction sequence is influenced by or limited by financial constraints.}

\subsection{Classification of definitions}
The definitions of dPISiE, PISiE, and ISE are such that these allow a person to construct a mental picture of the defined concept from the definition alone. That is no prior knowledge or intuition about the concept under definition is required
as a precondition for developing the mental picture from the definition. We call a definition with that property a {\em constructing definition}.%
\footnote{As an alternative to  ``constructing definitions'' we suggest ``imaginative definition'' (or ``imaginational definition'').} Of course after having met these definitions, the second and later time one takes notice of them the constructing of a mental picture does not take place anymore because it is in place already. The further role of a constructing definition, viewed from the perspective of a user or reader who meets the definition several times in succession, is to stabilize or to reinforce a mental picture rather than to construct it. 

Besides constructing definition an ostentative definition can be considered. An ostentative definition of 
some object class $C$ will allow one to point at instances of $C$ in a context where members of other classes are around as well. For instance, given the notion of a mammal, one may want to define the class of elephants. This can be achieved by an ostentative definition. In some cases the ostentative definition is merely a listing of features from a very specific kind of object, for instance a technological artifact, which is informally given in advance. Then the definition is an abstracting definition. One abstracts from the known details of a complex (kind of) entity. An abstracting definition of ISE is found if one takes in mind a specific processor architecture and provides a description of that at some level of abstraction. Abstracting definitions are constructing a mental picture to a lesser extent than constructing ones.

Given the ongoing change of technology abstracting definitions based on specific artifacts will become outdated, the
consequence being that new definitions must be sought. This leads us to the notion of a spiraling definition, or alternatively, a changing perspective definition, or in a more informal wording, a moving target definition. In the case of a spiraling definition each definition of a concept merely provides a platform from which a better view of the matter can be developed and from which a new and improved definition can be sought.

Spiraling definitions appear in difficult cases such as life, freedom, democracy, and  sustainability.

\section{Conclusion}
Building upon the notions of putting into effect and directly putting into effect the notion of instruction sequence execution has been provided with a result sandwich definition. Further development of the definition brings one into the realm of spiraling definitions. In a spiraling definition it is known in advance that the effort of giving a definition may at best lead to a next level of understanding from which matters look potentially different and which provides novel incentives for improving the definition.

Perhaps computer technology is evolving in such a way that customer developed instruction sequences are most likely not executed by
any machine on the market. In that scenario, within 
experimental settings the execution of an instruction sequence may still be achieved, whereas in normal applications
instruction sequence execution may lead to security risks and performance problems which are definitely to 
be avoided. Java programming and its virtual machine technology constitute a clear step in that direction.

Once execution is given up as an objective for control code, and interpretation has gained primacy, the notion of an instruction sequence
itself becomes less important as well. It seems fair to say that instruction sequences are especially useful for expressing algorithms that
are supposed to be executed. Once the objective of execution has been abandoned higher level control code notations 
are likely to attract more computer user attention than instruction sequence notations can attract.

\end{document}